\documentclass[twocolumn, trackchanges]{aastex701}

\begin{document}

\title{Modeling the Accretion of High-Velocity Clouds from a Rotating Halo}

\author[orcid=0009-0003-9296-3161,gname='Izumi', sname='Seno']{Izumi Seno}
\affiliation{Department of Physics, Graduate School of Science, Nagoya University, Furo-cho, Chikusa-ku, Nagoya, Aichi 464-8602, Japan}
\email[show]{seno.izumi.y0@nagoya-u.jp}  

\author[orcid=0000-0003-4366-6518,gname='Shu-ichiro', sname='Inutsuka']{Shu-ichiro Inutsuka}
\affiliation{Department of Physics, Graduate School of Science, Nagoya University, Furo-cho, Chikusa-ku, Nagoya, Aichi 464-8602, Japan}
\email{inutsuka.shu-ichiro.i2@f.mail.nagoya-u.ac.jp}  

\author[orcid=0000-0003-3383-2279, sname='Jiro', sname='Shimoda']{Jiro Shimoda} 
\affiliation{Institute for Cosmic Ray Research, The University of Tokyo, 5-1-5 Kashiwanoha, Kashiwa, Chiba 277-8582, Japan}
\email{jshimoda@icrr.u-tokyo.ac.jp}  

\begin{abstract}

High-Velocity Clouds (HVCs) are a major fuel reservoir for star formation in the Galactic disk. 
Determining their origin and kinematics is thus crucial for understanding Galactic evolution. 
In this paper, we employ simple test-particle simulations to model HVC kinematics, generating line-of-sight velocity maps and probability density functions (PDFs) for comparison with observational results. 
We find that models assuming low angular momentum and an initial scale of tens of kiloparsecs (kpc) successfully reproduce the observed kinematic trends for both blue-shifted and red-shifted components. 
This consistency may support the dominance of intermediate-halo dynamics (tens of kpc scale) in regulating Galactic evolution, consistent with HVC formation via thermal instability in metal-polluted gas in the halo. 
Furthermore, by considering the entire bulk mass involved in the continuous accretion process---including diffuse or ionized components that often escape direct observation---our theoretical estimates yield a total mass accretion rate of several $M_\odot\ {\rm yr{}^{-1}}$. This indicates that HVC accretion has the potential to supply a sufficient amount of gas to the Galactic disk to sustain ongoing star formation over several Gyr. 
Our findings suggest that the Galactic baryon cycle and disk evolution are governed by dynamics within the intermediate halo, providing key kinematic constraints for future magnetohydrodynamical simulations that resolve spatial structures of high velocity clouds.

\end{abstract}



\section{Introduction}\label{sec:Intro}
The Milky Way galaxy has sustained a remarkably steady star formation rate ($\mathrm{SFR}$) of approximately $\sim 1~M_\odot\ {\rm yr}^{-1}$ over the past $\sim 10~\mathrm{Gyr}$ \citep{Robitaille_Whitney2010, Chomiuk_Povich2011, Licquia_Newman2015, Haywood2014, Haywood+2016}. 
However, the total available cold gas mass in the Galactic disk (atomic HI and molecular gas) is only $\sim 10^9\ M_\odot$ \citep[e.g.,][]{Kalberla+2009}. 
At the current $\mathrm{SFR}$, this reservoir would be depleted in only about $1\ \mathrm{Gyr}$ if not replenished. 
This disparity highlights the need for a continuous gas supply from the outer region, a process known as gas accretion, which is essential for maintaining long-sustained, steady-state star formation and completing the Galactic baryon cycle \citep[e.g., ][]{Tumlinson+2017, Shimoda+2024}.

High-Velocity Clouds (HVCs) are widely considered a leading candidate for this gas supply. 
HVCs are observationally identified by their HI gas emission, but should be inherently multiphase systems and are defined kinematically by a line-of-sight velocity that deviates by more than $90\ \mathrm{km}\ \mathrm{s}^{-1}$ from the velocity of Galactic rotation \citep[for the classic definition, see][]{Wakker_Woerden1997, Putman+2012b}. 
Observations show that HVCs are ubiquitously distributed across the entire sky \citep[e.g., the all-sky survey results from][]{Wakker+2004, Peek+2007, Putman+2012b}. Revealing the origin and dynamics of HVCs is therefore crucial not only for understanding complex halo dynamics but also for reconstructing the star formation history of the Galactic disk.

The origin of HVCs, however, remains a central and highly debated topic. 
Competing theories generally fall into four categories: (1) \textit{Galactic fountain}, recycling gas from the disk \citep[e.g.,][]{Shapiro_Field1976, Fraternali_Binney2006}; (2) \textit{Cold accretion}, involving gas streams from the IGM \citep[e.g.,][]{Keres+2005, Stewart+2011}; (3) \textit{Satellite stripping}, where gas is removed from satellite galaxies via tidal or ram pressure forces (e.g., the Magellanic Stream; \citealp{Fox+2014}); and (4) \textit{Thermal instability}, where gas condenses from the hot halo \citep[e.g.,][]{Field1965, Seno+2025}.
The primary obstacle to conclusively distinguishing between these theories has been the difficulty in accurately measuring the distance to HVCs using radio observations alone, which in turn prevents reliable estimation of their masses and proper motions \citep{Wakker_Woerden1997, Putman+2012b}.

Recently, progress has been made in constraining HVC distances using absorption line observations. 
By analyzing the spectra of background sources (such as Galactic halo stars or quasars) that lie behind HVCs, the presence or absence of HVC absorption lines can be used to determine the minimum or maximum distance to the HVC. While early studies often utilized optical lines (e.g., Ca II or Na I), the vast majority of these constraints have been achieved in the ultraviolet (UV) using Hubble Space Telescope spectroscopy \citep[e.g., ][]{Sembach+1999, Wakker+1999, Richter+2001, Thom+2006, Thom+2008, Wakker+2007, Shull+2009,Lehner_Howk2011, Richter+2017,Fox+2019}.
While reliable for single lines of sight, these pencil-beam measurements cannot determine the overall mass or spatial distribution of HVCs. 
Specifically, constraining the angular momentum of HVCs is crucial for determining whether they can effectively fuel the Galactic disk.

In this paper, we employ a simplified model for the motion of falling dense clouds under the Galactic potential to examine the overall motion and characteristic origin scale of HVCs. 
By modeling particles originating from various initial heights and angular momenta, we construct line-of-sight velocity maps and probability density functions. 
We then rigorously compare these kinematic results with observational catalogs to constrain the most viable spatial origin of the observed HVC population. 
Furthermore, we provide a physical estimate of the mass accretion rate to quantify the HVC contribution to the Milky Way's $\mathrm{SFR}$.


\section{Models}\label{sec:Model}
\citealp{Seno+2025} demonstrated that thermal instability in the Galactic halo produces cold clouds that condense to high densities. Given that these clouds are significantly denser than the ambient medium, we model their dynamics as test particles falling under the Galactic potential for simplicity.  
The equations of motion in cylindrical coordinates ($R, \phi, z$) are given by:
\begin{align}
     \dfrac{dv_R}{dt} &= -g_R + \dfrac{v_\phi^2}{R}, \label{eq:model:EoM_R} \\
     \dfrac{dv_z}{dt} &= -g_z,\label{eq:model:EoM_z}
\end{align}
where $g_R$ and $g_z$ are the radial and vertical components of the total gravitational acceleration, respectively. 
We use angular momentum conservation law in the $\phi$-direction, which allows us to determine the angular velocity $v_\phi$ at any time $t$ by $v_\phi (t) = R_0 v_{\phi, 0}/R(t)$, where $R_0$ and $v_{\phi, 0}$ are the initial radial position and angular velocity of the particle, respectively.

\subsection{Gravitational Acceleration}
\label{subsec:Model_g}

The total gravitational acceleration is derived from the potential $\Phi = \Phi_{\rm BD} + \Phi_{\rm HL}$, which includes contributions from the stellar components (bulge and disk) and the dark matter halo.
We adopt the axisymmetric Miyamoto-Nagai model for the gravitational potential ($\Phi_{\rm BD}$) formed by the stellar components of the Milky Way Galaxy (\citealp{Miyamoto_Nagai1975}):
\begin{align}
     \Phi_{\rm BD} (R,z) = - \sum_{i = 1}^{2} \frac{GM_i}{\sqrt{R^2 + (a_i + \sqrt{z^2 + b_i^2})^2}}.
     \label{eq:model:gravity:BD}
\end{align}
Here, $G$ is the gravitational constant. 
The parameters for the bulge ($i=1$) and disk ($i=2$) are set as $a_i = (0; 7.258) \mathrm{~kpc}$, $b_i = (0.495; 0.520) \mathrm{~kpc}$, and $M_i = (2.05 \times 10^{10}; 2.547 \times 10^{11})~M_\odot$, respectively.

For the dark matter (DM) halo, we assume a Navarro-Frenk-White (NFW)-like density profile (\citealp{Navarro_etal1996}):
\begin{align}
     \varrho = \dfrac{\varrho_0}{x ( 1 + x)^2} - \dfrac{\varrho_0}{x_v ( 1 + x_v)^2},
     \label{eq:model:gravity:DM}
\end{align}
where $\varrho$ is the DM density and $x \equiv r / r_c$ is the radius normalized by the core radius $r_c$. The parameters $\varrho_0$ and $x_v \equiv r_v / r_c$ characterize the total mass and size of the DM halo. The corresponding DM potential ($\Phi_{\rm HL}$) is derived by solving the Poisson's equation:
\begin{align}
     \dfrac{1}{r^2} \dfrac{d}{dr} \left ( r^2 \dfrac{d\Phi_{\rm HL}}{dr} \right ) = 4 \pi G \varrho , 
     \label{eq:model:gravity_DM_potential}
\end{align}
which yields the radial derivative of the potential:
\begin{align}
     \dfrac{d\Phi_{\rm HL}}{dr} = 
     \left \{
          \begin{array}{ll} 
               4\pi G \varrho_0 r_c 
               \left [ 
                    - \dfrac{1}{x(1+x)} \right. \\ [3mm]
                    \left. {} + \dfrac{\ln (1 + x)}{x^2} - \dfrac{x}{3x_v(1+x_v^2)}\right ], & ( x \le x_v ) 
                    \\ [5mm]
               \dfrac{G M_{\rm HL}}{r_c^2 x^2}, & ( x > x_v ) 
          \end{array} 
     \right .
\end{align}
The total mass of the dark matter halo ($M_{\rm HL}$) is given by:
\begin{align}
     M_{\rm HL} = 4\pi r_c \varrho_0 \left [ \ln (1 + x_v) - \dfrac{x_v}{1 + x_v} - \dfrac{x_v^2}{3 (1 + x_v)^2} \right ]. 
     \label{eq:mode:gravity:TotMass_DM}
\end{align}
For our calculations, we set the parameters to $\varrho_0 = 1.06 \times 10^7 M_\odot \mathrm{~kpc}^{-3}$, $M_{\rm HL} = 10^{12} M_\odot$, and $r_v = 300 \mathrm{~kpc}$ (\citealp{Sofue2012}). 
These values uniquely determine the core radius as $r_c \approx 15.408 \mathrm{~kpc}$.

The total radial ($g_R$) and vertical ($g_z$) gravitational accelerations ($\mathbf{g} = -g_R \mathbf{e}_R - g_z\mathbf{e}_z$) created by all Galactic components are derived from the total potential $\Phi = \Phi_{\rm BD} + \Phi_{\rm HL}$ as:
\begin{align}
     g_R &= \dfrac{\partial\Phi_{\rm BD}}{\partial R} + \dfrac{R}{\sqrt{R^2 + z^2}}\dfrac{d\Phi_{\rm HL}}{dr},
     \\
     g_z &= \dfrac{\partial\Phi_{\rm BD}}{\partial z} + \dfrac{z}{\sqrt{R^2 + z^2}}\dfrac{d\Phi_{\rm HL}}{dr}.
\end{align}

\subsection{Parameters of Initial Conditions}
\label{subsec:Model_param}
\begin{table}[t]
     \centering
     \caption{Parameters of the initial conditions of our models.}\label{tab:model:param}
     \begin{tabular}{lcc}\hline
          Model     &    $z_{\rm max}$ [kpc]     &    $f_{\rm rot}$ \footnote{The parameter $f_{\rm rot}$ is the ratio of the initial angular velocity to the local circular velocity of the Galactic disk at the radius $R_0$ that corresponds to the initial position of the cloud $(R_0, z_0)$.}  \\  \hline \hline  
          z3R0.1  &    3    &    0.1  \\
          z20R0.1 &    20   &    0.1  \\
          z20R1.0 &    20   &    1    \\
          z20R2.0 &    20   &    2    \\
          z300R0.1&    300  &    0.1  \\
          Outflow\footnote{This is the outflow model. Elements are initialized at the disk plane ($z_0 = 0.1\ \mathrm{kpc}$) with an initial vertical velocity of $v_{z, 0} = f_{\rm out} \sqrt{2\Phi (R_0, z_0)}$, where $f_{\rm out}$ is the fraction of the initial velocity relative to the escape velocity ($0 < f_{\rm out} < 1$).}      
                    &    0.1    &    1    \\
          \hline
     \end{tabular}
\end{table}

To investigate the effects of various inflowing motions on the Galactic disk, we conduct a series of calculations by systematically varying the initial conditions of the test particles. 
The key parameters defining our models are summarized in Table \ref{tab:model:param}.
The initial positions of the test particles in the radial ($R$), azimuthal ($\phi$), and vertical ($z$) directions were set as follows: 
     the radial range is $1\ \mathrm{kpc} \le R_0 \le 10\ \mathrm{kpc}$ with a uniform spacing of $\Delta R = 1\ \mathrm{kpc}$; 
     the azimuthal angle covers the full range ($0 \le \phi_0 \le 2\pi$) with an interval of \(\Delta \phi = \pi / 4\); 
     and the vertical range is $1\ \mathrm{kpc} \le z_0 \le z_{\rm max}\ \mathrm{kpc}$ with a spacing of $\Delta z = 1\ \mathrm{kpc}$.

We systematically explore two primary parameters across our models. 
First, the upper limit of the initial vertical height ($z_0$) defines the maximum vertical distance (in kpc) from the Galactic plane at which the accreting elements are initially distributed. 
We test initial heights of $z_{\rm max} = 3\ \mathrm{kpc}$ (representing the inner halo), $z_{\rm max} = 20\ \mathrm{kpc}$ (the intermediate halo), and $z_{\rm max} = 300\ \mathrm{kpc}$ (approximating the virial radius).
Elements start in free-fall with initial velocity $\bm{v}_0 = (0, v_{\phi, 0}, 0)$. 
Second, the initial angular velocity ($v_{\phi, 0}$) is determined by the rotational velocity ratio ($f_{\rm rot}$) relative to the local circular velocity of the Galactic disk ($V_{\rm rot}$). 
This local velocity is defined for the initial positions as $V_{\rm rot} = \sqrt{g_R(R_0, z_0) R_0}$, yielding $v_{\phi, 0} = f_{\rm rot}V_{\rm rot}$. 
We examine three cases: $f_{\rm rot}=0.1$ (low angular momentum), $f_{\rm rot}=1$ (co-rotating), and $f_{\rm rot}=2$ (super-rotating). 
The model names, such as $\rm{z3R0.1}$, are constructed to clearly reflect the corresponding values of $z_0$ and $f_{\rm rot}$. 

Additionally, the Outflow model is included as a comparison case for gas outflow rather than inflow. 
For this model, particles are initialized at the disk plane ($z_0 = 0.1\ \mathrm{kpc}$) with an initial velocity of $\bm{v}_0 = (0, V_{\rm rot}, v_{z, 0})$. 
In the Outflow model, the initial vertical velocity ($v_{z, 0}$) is defined as a fraction of the escape velocity at the element's initial position ($R_0$): $v_{z, 0} = f_{\rm out} \sqrt{2\Phi (R_0, z_0)}$. 
Here, $f_{\rm out}$ represents the fraction of the initial vertical velocity relative to the escape velocity. 
To simulate the varying conditions typical of a Galactic fountain scenario\citep[e.g.,][]{Shapiro_Field1976, Bregman1980, Fraternali_Binney2006}, we randomly set $f_{\rm out}$ within the range $0 < f_{\rm out} < 1$.

\section{Results}\label{sec:Results}
In this section, we present the results derived from the model setup detailed in Section \ref{sec:Model}.
First, we generate all-sky maps of the line-of-sight velocity (Section \ref{subsec:Results:los}) to compare the kinematic signatures produced by different rotational models. 
Next, we construct the probability density function (PDF) of the modeled HVCs in Section \ref{subsec:Results:pdf}. 
Finally, in Section \ref{subsec:Results:acc}, we utilize our calculations to provide a brief estimate of the gas accretion rate onto the Galactic disk.

\subsection{All-Sky Map of Line-of-Sight Velocity}
\label{subsec:Results:los}

\begin{figure*}
     \centering
     \includegraphics[width = \linewidth]{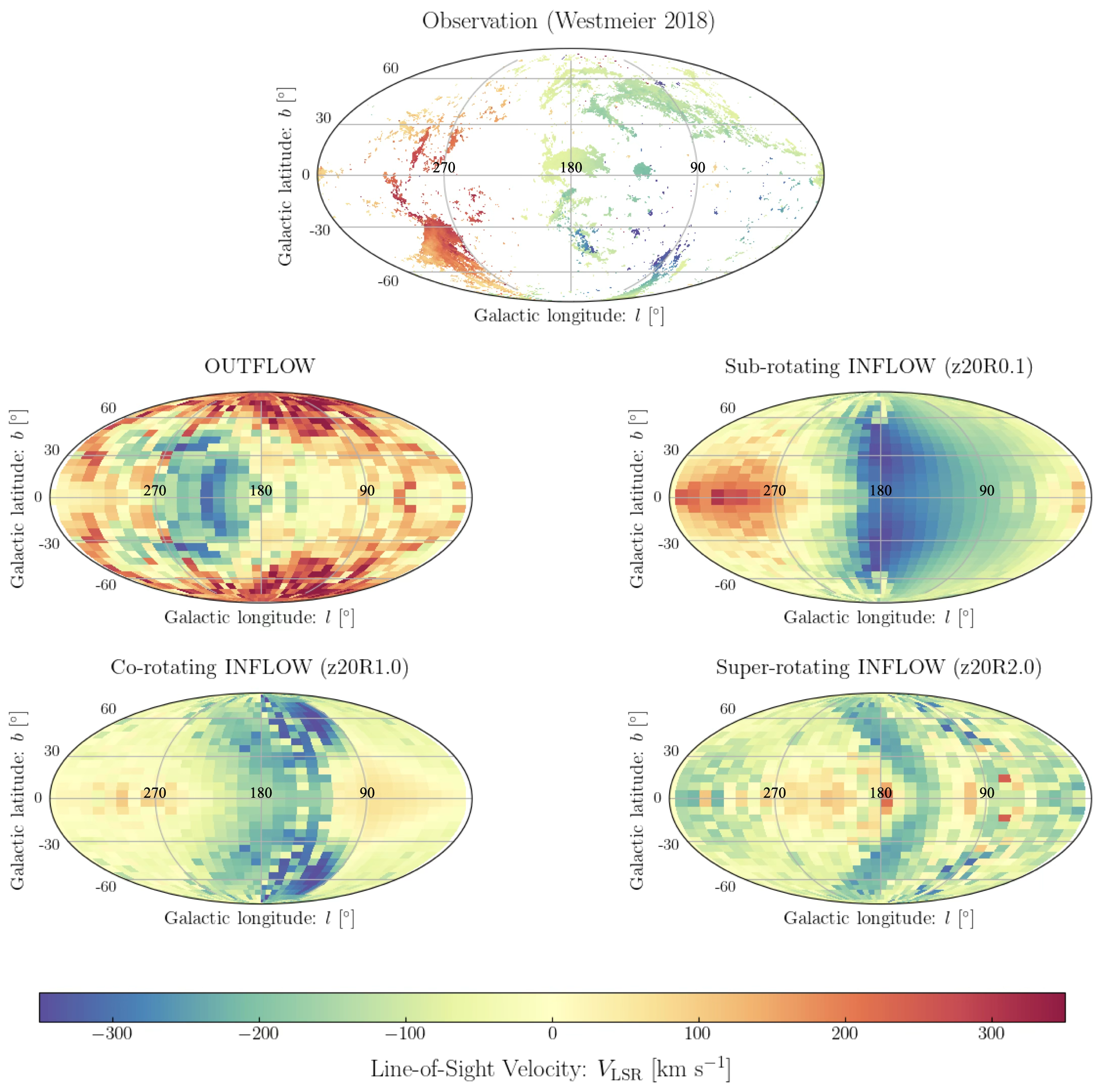}
     \caption{$V_{\rm LSR}$ map in Galactic coordinates, where $l$ and $b$ are the Galactic longitude and latitude, respectively. 
     The color shows $V_{\rm LSR}$. 
     This figure compares the observation (top panel) and four distinct models (bottom panels). Top panel: Observational all-sky map constructed from the catalog \citep{Westmeier2018}, based on the HI4PI Survey \citep{HI4PI+2016}. Bottom panels: Comparison of four distinct models: (top-left) the Outflow model, and three inflow models with varying rotational ratios: (top-right) sub-rotating inflow (`z20R0.1'), (bottom-left) co-rotating inflow (`z20R1.0'), and (bottom-right) super-rotating inflow (`z20R2.0').}
     \label{fig:LSRmap}
\end{figure*}

First, we generate an all-sky map of the line-of-sight velocity in the local standard of rest (LSR) frame ($V_{\rm LSR}$). 
Figure \ref{fig:LSRmap} shows the resulting $V_{\rm LSR}$ map in Galactic coordinates, where $l$ and $b$ are the Galactic longitude and latitude, respectively. 
For comparison, the top panel displays the observational all-sky map constructed from the catalog \citep{Westmeier2018}, based on the HI4PI Survey \citep{HI4PI+2016}.
The four bottom panels show the calculation results for four distinct models: the Outflow (fountain) model (`OUTFLOW', top-left) and three inflow models characterized by varying rotational ratios. 
Specifically, the sub-rotation model ('z20R0.1', top-right) exhibits large-scale kinematic features that are broadly consistent with the global velocity gradients in the observed all-sky map, while other models, including the Outflow model, do not.
The Outflow model, for instance, shows that, unlike the inflow models, the red-shifted component ($V_{\rm LSR} > 0$) dominates at high Galactic latitudes ($b \gtrapprox 45^{\circ}$), which is inconsistent with the observed coherent large-scale kinematics \citep{Putman+2012b,Westmeier2018}.

The sub-rotation model (z20R0.1) and observations share two key similarities. 
We exclude the region around the Large and Small Magellanic Clouds (LMC, SMC) and the Magellanic Stream (roughly $180^\circ < l < 360^\circ$, $b \sim -45^\circ$) to avoid contamination from these local structures neglected in our axisymmetric models.
     First, elements at low Galactic latitudes ($|b| \lessapprox 45^{\circ}$) show a systematic variation in $V_{\rm LSR}$ across Galactic longitudes ($l$). 
     Second, at high Galactic latitudes ($|b| \gtrapprox 45^{\circ}$), the frequency of red-shifted LSR velocities ($V_{\rm LSR} > 0$) significantly decreases, with blue-shifted LSR velocities ($V_{\rm LSR} < 0$) becoming strongly dominant.
The first trend is kinematic evidence of low angular momentum. 
The inflowing gas rotates more slowly than the Galactic disk at the LSR. 
Since these elements are observed along the disk's rotational motion ($\approx 220\ \mathrm{km~s^{-1}}$), the rotational deficit causes elements at Galactic longitudes $l \lessapprox 180^{\circ}$ to appear to move towards the observer, and conversely, elements at $l \gtrapprox 180^{\circ}$ to appear to move away from the observer. 
This characteristic pattern signifies gas possessing low angular momentum relative to the disk.

     Notably, the sub-rotational kinematics predicted by the z20R0.1 model appears to be consistent with the observed velocity of the Leading Arm \citep[e.g.,][]{DOnghia_Fox2016} located at $l \approx 270^\circ$ and $b \lessapprox 30^\circ$, suggesting that our model can naturally account for its large-scale kinematic features.

The second trend—the prevalence of $V_{\rm LSR} \lesssim 0$ at high latitudes—is consistent with infalling motion. 
At high $b$, the line-of-sight velocity is dominated by the vertical velocity component ($v_z$), allowing us to directly observe the continuous inflowing motion (i.e., $v_z < 0$).
The general agreement between our low-angular-momentum accretion model (`z20R0.1') and the observational results \citep{Putman+2012b,Westmeier2018} suggests that the overall motion of HVCs currently observed in the Milky Way halo is consistent with ongoing accretion. 

Our current calculation simplifies the system by neglecting angular momentum transport mechanisms; 
     thus, a more detailed understanding of the long-term evolution and large-scale Galactic outflow requires more sophisticated numerical simulations that include fluid dynamics. 
Nonetheless, the fact that inflowing motion is kinematically favored implies that gas accretion is not significantly hindered by outflows. 
This suggests that the hydrostatic equilibrium state, often used as an initial condition \citep[e.g., ][]{Seno+2025}, provides a reasonable setup for investigating the formation and subsequent accretion of HVCs.

Note that our test-particle model inherently produces a smooth velocity field, whereas the actual HVC data show a highly clumpy distribution with small-scale structure \citep[e.g.,][]{Clark+2022}. 
Since our simplified approach neglects hydrodynamical instabilities, thermal conduction, and magnetic fields, it cannot capture the detailed properties of the observed HVCs. 
In this letter we focus only on the global kinematics rather than to reproduce the detailed properties of the observed HVCs.

\subsection{Probability Density of HI HVCs}
\label{subsec:Results:pdf}
\begin{figure*}
     \centering
     \includegraphics[width = \linewidth]{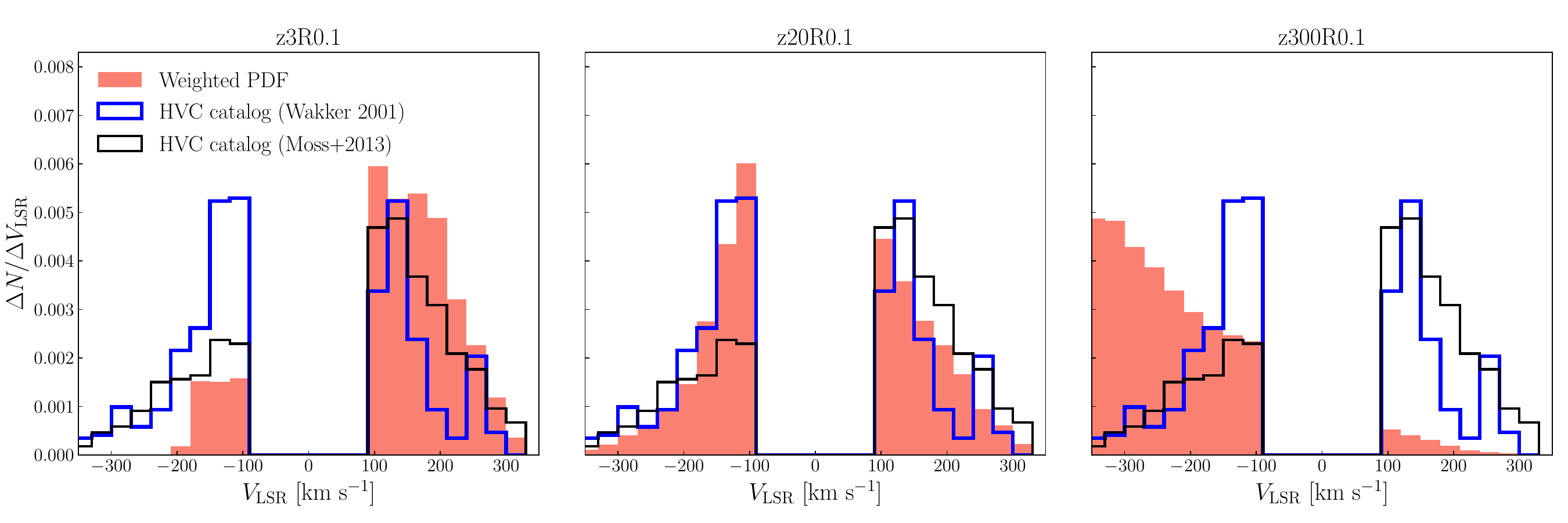}
     \caption{The probability density function (PDF) of $V_{\rm LSR}$ weighted by the inverse of the distance from the observer, as indicated in Equation \eqref{eq:Results:pdf}.
     The red histogram is derived from our calculation, while the blue- and black-histograms are derived from the HVCs catalogs in \cite{Wakker2001} and \cite{Moss+2013}, respectively.
     This figure compares three distinct sub-rotation models ($f_{\rm rot} = 0.1$): 
     (left) inner-halo (`z3R0.1'), (middle) intermediate-halo (`z20R0.1'), and (right) whole-halo (`z300R0.1').}
     \label{fig:LSRhistgram}
\end{figure*}

Next, we discuss the typical kinematic properties of HVCs by analyzing the distribution of $V_{\rm LSR}$ derived from our calculations. 
To account for the observational bias toward closer object in easier detection, we construct the probability density function (PDF, $\Delta N/\Delta V_{\rm LSR}$) weighted by the inverse of the squared distance from the observer ($D$):
\begin{align}
     \dfrac{\Delta N}{\Delta V_{\rm LSR}} \equiv \dfrac{\sum_{i \in I(V_{\rm LSR}, V_{\rm LSR} + \Delta V_{\rm LSR})} 1 / D_i^2}{\sum_i 1 / D_i^2},
     \label{eq:Results:pdf}
\end{align}
where $i$ indicates the index of the observed element at time $t$.
We introduce the above weight by the following consideration: 
The spatial resolution of these observation is $\sim 16^\prime$. 
If the HI clouds are located further than the currently observed cloud, their apparent size on the plane of the sky can be less than the observational beam size. 
As a result, distant clouds are more susceptible to beam dilution ($\propto 1/D^2$) and are statistically more likely to be missed in observational surveys.

Figure \ref{fig:LSRhistgram} shows the weighted PDF of $V_{\rm LSR}$. 
The red histogram represents the distribution derived from our calculation using Equation \eqref{eq:Results:pdf}. 
Note that we only include data from HVCs (defined by the typically high $|V_{\rm LSR}|$) in these PDFs, as it is challenging to reliably distinguish smaller $V_{\rm LSR}$ components from the high velocity dispersion of the Galactic disk gas.

By comparing our results with observational catalogs (blue: \citealp{Wakker2001}; black: \citealp{Moss+2013}), the intermediate-halo model (z20R0.1) shows the most consistent result. 
Both the observational results and the intermediate-halo model demonstrate that blue-shifted ($V_{\rm LSR} < 0$) and red-shifted ($V_{\rm LSR} > 0$) HVCs exist in roughly equal proportions. 
Conversely, the inner-halo model (z3R0.1) significantly underestimates the fraction of blue-shifted elements, while the whole-halo model (z300R0.1) overestimates it.

This consistency with the intermediate-halo model suggests that HVCs in the Milky Way are primarily distributed within tens of kpc from the Galactic mid-plane, supporting a picture where the gas originates from the intermediate halo. 
The tens-of-kpc scale identified here offers a distinct perspective on accretion. 
This intermediate scale contrasts with both cosmological cold accretion from virial radii ($\sim 100$ kpc; e.g., \citealp{Fraternali_Binney2006}) and classical galactic fountains restricted to a few kpc \citep[e.g.,][]{Shapiro_Field1976}. 
Modern fountain models incorporating cosmic-ray heating suggest that outflows can condense and free-fall from $z \simeq 6$--$8$ kpc. However, such fallback is confined to only $R < 5$ kpc \citep{Shimoda_Asano2024}.

Although our current model neglects angular momentum transport and requires further verification via hydrodynamic simulations, we note that thermal instability at this tens-of-kpc scale is theoretically supported by linear analysis of gravitationally stratified media \citep{Seno+2025}.

\subsection{Characteristic Size and Mass of HVCs}
\label{subsec:Results:acc}

\begin{figure}[htbp]
    \centering
    \includegraphics[width=0.49\textwidth]{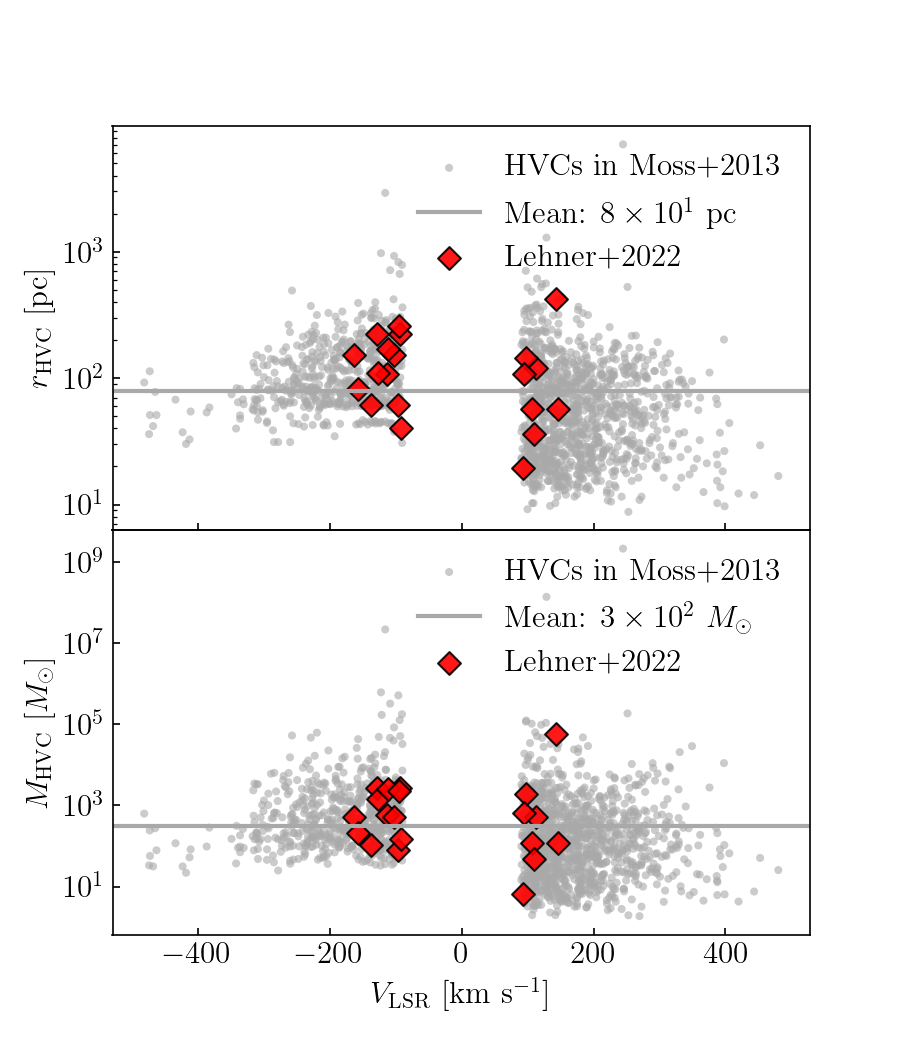}
    \caption{Estimated physical sizes (top panel) and masses (bottom panel) of HVCs as a function of the line-of-sight velocity ($V_{\rm LSR}$). 
    Grey dots represent the observational data from the catalog of \citet{Moss+2013}, where distances were assigned via a nearest-neighbor search in our simulation's phase space. 
    The grey solid lines indicate the mean values of the sizes and masses. 
    Red dots denote HVCs with distance constraints from absorption-line studies using Gaia distances \citep{Lehner+2022}. }
    \label{fig:Results:size_mass}
\end{figure}

Finally, we estimate the characteristic size and mass of HVCs based on our modeling. 
Since our simulation provides the distance from the observer ($D$) in addition to the coordinates $(l, b, V_{\rm LSR})$, we can resolve the distance ambiguity of observed HVCs by performing a nearest-neighbor search in the 3D phase space of $(l, b, V_{\rm LSR})$. 
Figure \ref{fig:Results:size_mass} shows the estimated sizes and masses of HVCs as a function of $V_{\rm LSR}$. 
The grey dots are derived from the HVC catalog by \citet{Moss+2013}, where the physical size is estimated as $r_{\rm HVC} \sim D\sqrt{\Omega}$ (with $\Omega\ [{\rm deg^2}]$ being the angular extent) and the mass as $M_{\rm HVC} \sim N_{\rm HI} r_{\rm HVC}^2$ ($N_{\rm HI}$ is the column density). 
For further comparison, the red dots represent HVCs from \citet{Lehner+2022}, where distances were constrained via absorption lines in halo stars with Gaia distances \citep{Gaia2016, Gaia2018, Gaia2021}.
According to the nearest-neighbor search, the characteristic size of HVCs is estimated to be $\simeq 80\ \mathrm{pc}$ and the typical mass is estimated to be $\simeq 3 \times 10^{2}\ M_\odot$.

\section{Discussion}\label{sec:Discussion}

As demonstrated in Section \ref{sec:Results}, the kinematic properties of the HVC population currently observed in the Milky Way halo are best reproduced by models where clouds originate from thermal instability within the intermediate halo (a few tens of kpc).
Let us adopt this scenario and try to theoretically estimate the characteristic physical properties of newly formed HVCs.
Under this hypothesis, the characteristic size is determined by the Field length ($\lambda_{\rm F}$):
\begin{align}
     \lambda_{\rm F} \equiv \sqrt{\dfrac{\mathcal{K}\langle T \rangle}{\langle n \rangle^2 \Lambda}} 
      =\ & 1.2\ {\rm kpc} 
     \left ( \dfrac{\langle n \rangle}{10^{-3}\ {\rm cm^{-3}}}\right )^{-1}
     \left ( \dfrac{\langle T \rangle}{10^6\ {\rm K}}\right )^{7/4} 
     \notag\\
     & \times \left ( \dfrac{\Lambda}{10^{-22}\ {\rm erg~s^{-1}cm^{3}}}\right )^{-1/2},
     \label{eq:Results:acc_size}
\end{align}
where $\langle n \rangle$ and $\langle T \rangle$ represent the mean number density and temperature of the ambient halo gas, and $\Lambda$ is the cooling function. 
We adopt the thermal diffusivity for a fully ionized plasma, $\mathcal{K} = 1.2 \times 10^{-6}\ T^{5/2}\ \mathrm{erg~s}^{-1}\mathrm{cm}^{-1}\mathrm{K}^{-1}$ \citep{Parker1953,Spitzer1962}. 
For the typical circumgalactic medium (CGM), we assume baseline parameters of $\langle n \rangle \sim 10^{-3}\ {\rm cm}^{-3}$ and $\langle T \rangle \sim 10^6\ {\rm K}$, consistent with X-ray observations estimating a total hot gas mass of $\sim 10^{11}\ M_\odot$ in the Milky Way halo \citep{Fang+2013, Miller_Bregman2013, Miller_Bregman2015, Nakashima+2018}. 
Assuming a spherical geometry, the initial mass of an HVC ($M_{\rm HVC}$) formed at this scale is given by:
\begin{align}
     M_{\rm HVC} &\equiv \dfrac{4\pi}{3} \rho \lambda_{\rm F}^3,
     \label{eq:Results:acc_mass}\\
     &\approx 8.7 \times 10^4\ M_\odot
     \left ( \dfrac{\langle n \rangle}{10^{-3}\ {\rm cm^{-3}}}\right )
     \left ( \dfrac{\lambda_{\rm F}}{1.2\ {\rm kpc}}\right )^3.
     \notag
\end{align}

The typical sizes and masses derived from Milky Way observations are one to two orders of magnitude smaller than these theoretical prediction if we adopt $\langle n \rangle \sim 10^{-3}\ \mathrm{cm}^{-3}$, typical values of the density of the Galactic halo far away from the disk. 
In this section, we discuss three potential factors important in the estimation: observational bias (Section \ref{subsec:Discussion:bias}), the presence of ionized components (Section \ref{subsec:Discussion:ionized}), and magnetic field effects (Section \ref{subsec:Discussion:mag}). 
Finally, in Section \ref{subsec:Discussion:acc_rate}, we integrate these factors to derive a realistic estimate of the mass accretion rate onto the Galactic disk.

\subsection{Observational Bias}
\label{subsec:Discussion:bias}

In contrast to the Milky Way, studies of HVCs associated with the face-on galaxy M83 \citep{Miller+2009, Nagata+2025} report typical sizes of $\sim 1\ {\rm kpc}$ and masses of $\sim 10^{5}\ M_\odot$, values that match our theoretical predictions for the diffuse halo.
The difference between Galactic values and extra-galactic values might be explained by a combination of the Galactic spatial density profile and observational bias. 
In the case of M83, the system is viewed externally, whereas Milky Way HVCs are observed from the position within the disk, i.e., the Solar System's position near the Galactic mid-plane. 
Since gas density decreases with height ($z$) above the Galactic plane, HVCs forming closer to the disk originate in environments with higher ambient densities. 
Given that the Field length and cooling time scale as $r_{\rm HVC} \propto n^{-1}$ and $\tau_{\rm cool} \propto n^{-1}$, respectively, higher-density regions naturally produce smaller, more rapidly cooling clouds. 
Since the cloud mass scales as $M_{\rm HVC} \sim \rho \lambda_{\rm F}^3 \propto n^{-2}$, an increase in density by an order of magnitude results in a decrease in mass by two orders of magnitude.

Re-evaluating our model with a typical density of $\langle n \rangle = 10^{-2}\ {\rm cm}^{-3}$, which characterizes regions closer to the disk, yields $r_{\rm HVC} \sim 100\ {\rm pc}$ and $M_{\rm HVC} \sim 10^{3}\ M_\odot$. 
As shown in Figure \ref{fig:Results:size_mass}, these values align well with observational data from the Milky Way. 
Because nearby HVCs are more easily detected, the population observed within our Galaxy is likely dominated by clouds formed in the higher-density regions near the disk. 
Conversely, in M83, HVCs at higher altitudes (where densities are lower) are more readily observed. Their larger sizes may obscure smaller clouds in the lower layers, leading to a bias toward larger observed masses and sizes.

\subsection{Ionized Components}
\label{subsec:Discussion:ionized}

Another substantial factor is the massive ionized envelope associated with HVCs.
Recent observations using absorption-line spectroscopy \citep[e.g.,][]{Shull+2009, Lehner_Howk2011, Richter+2017, Fox+2019} suggest that a significant amount of ionized gas surrounds the HI-HVCs, dominating the mass budget relative to the neutral component.
Recent numerical simulations \citep[e.g.,][]{Lucchini+2025} also indicate that the ionized component contributes approximately six times more to the total HVC mass than the HI component.
In addition, they show that the ionized accretion rate could account for $\simeq 80$ \% of the galactic star formation rate ($\sim 1\ M_\odot {\rm yr}^{-1}$ in the Milky Way case), while the neutral accretion rate can only balance $\sim 11$ \%.

Moreover, if thermal instability is the primary formation mechanism, the HI core must be continuously connected to the ambient hot gas. 
In this view, the ionized envelope represents the cooling interface, behaving similarly to the transition front between cold- and warm-neutral phases in the interstellar medium \citep{Nagashima+2005,Nagashima+2006}.
The thickness of this transition layer is governed by the Field length ($\lambda_{\rm F}$) of the hotter component (Equation \ref{eq:Results:acc_size}). 
Since this scale is determined by the ambient hot gas, we estimate that the ionized envelope accounts for approximately 90\% of the total mass of an HVC.

\subsection{Effect of Magnetic Field}
\label{subsec:Discussion:mag}

We also consider the influence of magnetic fields on HVC formation.
Classically, thermal conduction is suppressed in the direction perpendicular to magnetic field lines, effectively reducing the Field length.
The thermal conductivity perpendicular to the magnetic field is given by \citep{Braginskii1965}:
\begin{align}
    \mathcal{K}_{\perp} = \dfrac{\mathcal{K}_{\parallel}}{1 + \omega_{\rm c}^2 \tau_{ep}^2},
    \label{eq:Results:mag}
\end{align}
where $\omega_{\rm c} = eB/m_{\rm e}c$ is the cyclotron frequency, $\tau_{ep}$ is the electron-ion mean free time, and $\mathcal{K}_{\parallel}$ is the classical thermal conductivity parallel to the magnetic field \citep{Spitzer1962}.
For a dilute plasma such as the hot halo gas of the Milky Way, assuming a typical magnetic field strength of $B \sim 1\ \mu{\rm G}$, the Hall parameter reaches $\omega_{\rm c} \tau_{ep} \sim 10^{11}$.
Consequently, $\mathcal{K}_\perp$ is negligible compared to $\mathcal{K}_{\parallel}$.

Furthermore, magnetic fields can directly suppress the growth of thermal instability.
Future magnetohydrodynamic (MHD) simulations are therefore essential to understand how HVC sizes are regulated---particularly in the direction perpendicular to magnetic fields---by both the anisotropic suppression of thermal conduction and the magnetic stabilization of the instability itself.

\subsection{Determination of Mass Accretion Rate}
\label{subsec:Discussion:acc_rate}

In this section, we provide a realistic estimate of the mass accretion rate onto the Galactic disk, taking into acount the considerations from the previous subsections.
Based on the discussion in Section \ref{subsec:Discussion:bias}, if we assume a mean ambient density of $\langle n \rangle = 10^{-2}\ {\rm cm}^{-3}$ within the inner 20 kpc of the halo (the condition required to reproduce observed HVC properties), the mass accretion rate is estimated as:
\begin{align}
    \dot{M}_{\rm acc} &\equiv
    2 \times \dfrac{\pi R_{\rm disk}^2 z_{\rm max} \langle \rho \rangle}{\tau_{\rm acc}} f_{\rm HVC} 
    \label{eq:Results:acc_rate_1} \\
    &
    \begin{aligned}
     \simeq\ & 
     8.5\ M_\odot\ {\rm yr}^{-1}
     \left ( \dfrac{R_{\rm disk}}{10\ {\rm kpc}}\right )^2
     \left ( \dfrac{z_{\rm max}}{20\ {\rm kpc}}\right )
     \\
     & \times 
     \left ( \dfrac{\langle n \rangle}{10^{-2}\ {\rm cm}^{-3}}\right )
     \left ( \dfrac{\tau_{\rm acc}}{0.1\ {\rm Gyr}}\right )^{-1}
     \left ( \frac{f_{\rm HVC}}{0.5}\right ).
    \end{aligned}
    \notag
\end{align}
Here, $f_{\rm HVC}$ represents the mass fraction of the halo gas that condenses into HVCs.
\citet{Lehner_Howk2011} estimated the total mass of the multiphase HVCs (HI $+$ HII) within $d \lesssim 12\ {\rm kpc}$ to be $\sim 10^8\ M_\odot$.
On the other hand, X-ray observations indicate that the mass of the ambient hot halo gas within a comparable inner volume ($r \lesssim 12\ {\rm kpc}$) is also on the order of $10^8\ M_\odot$ \citep[e.g., ][]{Miller_Bregman2013, Miller_Bregman2015}.
The identification of these comparable masses suggests that the total baryonic mass in the lower halo ($|z| \lesssim 20\ {\rm kpc}$) is distributed roughly equally between the ambient hot medium and the condensed HVC phase. 
Consequently, adopting a mass fraction of $f_{\rm HVC} \approx 0.5$ serves as a fiducial value for our calculation.
The accretion timescale is estimated as $\tau_{\rm acc} \equiv z / v_{\rm esc} \sim 0.1\ {\rm Gyr}$ for $z = 20\ {\rm kpc}$ and a virial velocity of $v_{\rm esc} \approx 170\ {\rm km\ s^{-1}}$.
Note that $z_{\rm max} = 20\ {\rm kpc}$ represents the original height of HVCs at their birth, distinguishing it from their present distribution, which is observed primarily within $12\ {\rm kpc}$ \citep{Lehner_Howk2011}.

Our theoretical accretion rate derived in Equation \eqref{eq:Results:acc_rate_1} is approximately one order of magnitude larger than recent observational estimates \citep[e.g., ][]{Shull+2009, Lehner_Howk2011, Richter+2017, Fox+2019}.
For instance, \citealp{Lehner_Howk2011} reported an inflow rate of $0.8$--$1.4\ M_\odot\ \mathrm{yr}^{-1}$ based on ionized HVCs, and \citealp{Fox+2019} derived $0.53 \pm 0.23\ M_\odot\ \mathrm{yr}^{-1}$.
This difference arises because observations measure the accretion rate strictly based on currently detectable, distinct clouds. 
In contrast, our theoretical derivation accounts for the entire bulk mass involved in the continuous accretion process, naturally incorporating diffuse or ionized ``unseen'' components that effectively fuel the disk but may escape direct observation.

%
%

Depending on the typical density of the ambient medium and the mass fraction of the ionized component, HVC accretion has the potential to supply a sufficient amount of gas to the Galactic disk to sustain ongoing star formation, a picture that is also supported by recent numerical simulations \citep[e.g., ][]{Lucchini+2025}.
However, quantitatively determining the total mass accretion rate remains a challenge.
This largely stems from the ambiguity in defining the ``ionized component'' in both observational and theoretical contexts, as well as the uncertainty in estimating the mass fraction $f_{\rm HVC}$ of the accreting gas.
To resolve this issue, future work requires high-resolution magneto-hydrodynamical simulations that fully resolve the transition layer thickness ($\sim \lambda_{\rm F}$).
Such simulations are essential to properly determine $f_{\rm HVC}$, physically define the structure of the ``ionized envelope'', and accurately quantify its overall contribution to the total mass accretion rate.

\section{Conclusion}
\label{sec:summary}

In this paper, we investigated the kinematic characteristics of High-Velocity Clouds (HVCs) by modeling their motion as free-falling test particles within the Galactic gravitational potential and comparing the resulting line-of-sight velocities with observational data. 
Our primary finding is that the HVC line-of-sight velocity maps and probability density functions (PDFs) are best described by inflow models with low angular momentum (sub-rotation) and an original birth place of tens of kpc, corresponding to the intermediate halo. 
This kinematic consistency strongly suggests that the HVC population currently observed is accreting from the intermediate halo while rotating more slowly than the Galactic disk, enabling the gas to fuel the disk.

Furthermore, by considering the entire bulk mass involved in the continuous accretion process---including diffuse or highly ionized ``unseen'' components that may escape direct detection---our theoretical estimates yield a total mass accretion rate of several $M_\odot\ \mathrm{yr}^{-1}$.
This indicates that HVC accretion has the potential to supply a sufficient amount of gas to the Galactic disk to sustain ongoing star formation over the timescale comparable to the age of the Galaxy.

However, since cloud formation via thermal instability is expected in regions where the density locally reaches $\langle n \rangle \sim 10^{-2}\ \mathrm{cm}^{-3}$ to reproduce the compact sizes of observed HVCs, strong perturbations—such as galactic winds or supernova feedback—may be necessary to provide sufficient density enhancements in the diffuse halo gas. 
High-resolution magnetohydrodynamical simulations that fully resolve the transition layer will be essential to properly determine the mass fraction ($f_{\rm HVC}$) of the accreting gas and define the structure of the ionized envelope, which are required to accurately quantify the total mass accretion rate. 
Such detailed theoretical modeling and further observational studies are expected to provide a realistic picture of the Galactic baryon cycle and disk evolution.

\begin{acknowledgments}
This work was supported by JSPS KAKENHI, Grant Nos. 24KJ1302 and 25H00394.
We acknowledge the use of the all-sky map of Galactic high-velocity clouds presented by \citet{Westmeier2018}, based on data from the HI4PI Survey \citep{HI4PI+2016}.
\end{acknowledgments}

\bibliography{sample701}{}
\bibliographystyle{aasjournalv7}



\end{document}